\documentstyle[11pt,newpasp,twoside]{article}
\def\emphasize#1{{\sl#1\/}}
\def\arg#1{{\it#1\/}}
\let\prog=\arg

\def\edcomment#1{\iffalse\marginpar{\raggedright\sl#1\/}\else\relax\fi}
\reversemarginpar

\begin{document}
\title{ A Keck/HST Survey for Companions to Low-Luminosity Dwarfs}
\author{D.W. Koerner}
\affil{University of Pennsylvania, David Rittenhouse Laboratory, 
209 S. 33rd St. Philadelphia, PA 19104-6396}
\author{J. Davy Kirkpatrick}
\affil{ Infrared Processing and Analysis Center, MS 100-22, 
California Institute of Technology, Pasadena, CA 91125}
\author{I.N. Reid}
\affil{University of Pennsylvania, David Rittenhouse Laboratory, 
209 S. 33rd St. Philadelphia, PA 19104-6396}
\author{J. Gizis}
\affil{University of Massachusetts, Dept. of Physics and Astronomy, LGRT 532A,
Amherst, MA 01003}

\begin{abstract}
We present preliminary results of an imaging survey for companions 
to low-luminosity dwarfs with spectral types ranging from M7 to L9. 
A K-band study with the Near Infrared Camera (NIRC)  
at the Keck telescope 
discriminates against background sources by searching for common 
proper motion. A complimentary
HST/WFPC2 snapshot survey is better able to resolve close companions, 
but is not as sensitive at wide separations.
Preliminary results from the Keck/NIRC survey have yielded the detection 
of 3 binaries in a sample of 10 L dwarfs, including one 
previously identified in HST imaging (Mart\'{\i}n et al.\ 1999). 
All three have equal-component luminosities and physical separations
between 5 and 10 AU. This result leads us to speculate that
binary companions to L dwarfs are common, that similar-mass 
systems predominate, and that their distribution peaks at radial 
distances in accord both with M dwarf binaries and with the radial 
location of Jovian planets in our own solar system. To fully establish 
these conjectures, however, will require quantitative analysis of an
appropriate sample. To this end, we outline a Bayesian scheme to 
test models of the underlying companion distribution with our completed
imaging survey.

\end{abstract}

\section{Introduction}

Recent detections of planetary and brown dwarf companions to nearby stars
have fueled efforts to undertake a complete inventory of circumstellar bodies
(Mayor \& Queloz 1995; Nakajima et al.\ 1995; Marcy \& Butler 1996). 
It is thus likely that we stand at the beginning of an exciting era of 
astronomical discovery in which a gradually unfolding census promises to
provide key evidence for the modes of origin for planets and binary stars.
As part of this endeavor, we have undertaken 
a search for companions to recently discovered low-luminosity 
field dwarfs (Kirkpatrick et al. 1999; 2000). Our study will provide a first 
look at the binary companion rate for sub-stellar objects. Furthermore, it 
will yield a special opportunity for imaging giant Jovian planets, since 
sensitivity is enhanced in the reduced glare of faint dwarf primaries.

The local field detection rate of very low-luminosity dwarfs in infrared sky 
surveys suggests they comprise a sizeable population which is well represented
by an extension of the field-star mass function, $\Psi(M) \propto 
M^{-\alpha}$, with $ 1 < \alpha < 2$ (Reid et al.\ 1999). The occurrence 
frequency of multiplicity among these systems is completely unknown; it is an 
open question as to whether the distribution of their companions matches that 
of M dwarfs or bears the stamp of a different, sub-stellar formation 
mechanism.  Stellar companions are detected in approximately 35\% of M dwarf 
systems with a distribution peaking at a radius in the range $3-30$ AU 
(Fischer \& Marcy 1992; Henry \& McCarthy 1993; Reid \& Gizis 1997). Efforts 
to uncover the mass and radial distribution of extra-solar planets around M 
stars are just beginning to meet with success and have revealed super 
Jovian-mass planets within a few AU of their central stars, consistent with 
results for earlier spectral types (Marcy et al.\ 1998). The relationship of 
this population to that of binary companions and planetary systems like our 
own is a topic of current debate (Black 1997). The true answer will not be 
readily apparent until a more complete 
range of mass and orbital distances has been surveyed.

To date, very few multiple systems have been identified with L-dwarf components. Several L-dwarf secondaries have been discovered around nearby stars 
(Becklin \& Zuckerman 1988; Rebolo et al.\ 1998; Kirkpatrick et al.\ 2000).  Among a handful of known binary brown-dwarf systems (e.g., Basri \& Mart\'{\i}n 1997), only two have primary spectral types as late as L: 
2MASSW J0345 is a double-lined spectroscopic L dwarf system (Reid et al.\ 1999), and DENIS-P J1228 was shown to be double in HST imaging observations 
(Mart\'{\i}n et al.\ 1999). The latter is composed of equal-luminosity 
components with a projected separation of 0.275$''$ (5 AU at the 18 pc
distance of DENIS-P J1228). Here we report preliminary results from a Keck
near-infrared imaging survey of a large sample of low-luminosity dwarfs 
and outline a complementary study with Hubble Space Telescope.

\section{Survey Characteristics}

\subsection{Keck/NIRC Imaging Survey}

Our target sample is culled from the 2MASS and DENIS near-infrared sky 
surveys and consists of objects spectroscopically confirmed to be L dwarfs
together with a smaller sample of nearby very late M dwarfs. 
Survey parameters are plotted in Fig.\ 1, including sky coverage, spectral 
type, and range of distances. Imaging is carried  
out at the Keck I telescope with NIRC, a cryogenically-cooled near-infrared 
camera which incorporates a 256$\times$256 Indium-antimonide array at the 
f/25 focus in an optical framework which yields a 0.15$''$ plate scale and 
38$''$-square field of view (Matthews \& Soifer 1994). 
The survey is sensitive to companions brighter than $m_{\rm K}$~=~21 at 
separations greater than 1$''$ (5-50 AU in the sampled range of 
distances) within a $20''\times20''$ square aperture (out to 100-1000 AU), 
and is capable of detecting components with luminosity close to that of the 
primary ($m_{\rm K} \sim 13$) at $\sim$0.3$''$ separation.
At this level of sensitivity, several additional sources are detected in a 
typical frame. Repeat observations in a second epoch, 
one year or more later, are being taken to determine if any of these share a 
common proper motion with the target; second-epoch observations are complete 
for only a subset of the sample which includes 10 L dwarfs at present. 

\begin{figure}
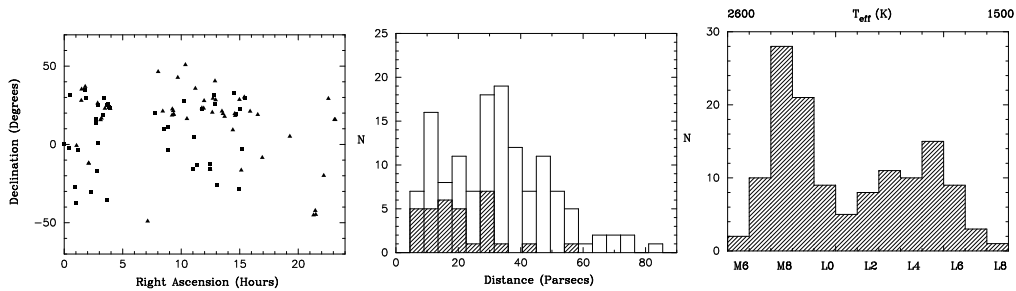

\plotfiddle{fig1a.ps}{0pt}{270}{20} {20} {-200}{0}
\plotfiddle{fig1b.ps}{0pt}{270}{20} {20} {-75}{25}
\plotfiddle{fig1c.ps}{0pt}{270}{20} {20} {50}{50}
\vskip 1.0truein
\caption{{\bf Left)} Sky coverage for Keck/NIRC survey of low-luminosity
dwarfs. Sample is culled from optical and near-infrared photometric
surveys in sky regions avoiding the galactic plane. L~dwarfs were identified
with follow-up spectroscopy. {\bf Center)} Range of distances
for objects in survey. Hatched bars represent those objects for which
distance has been determined by trigonometric parallax.  {\bf Right)} 
Distribution of spectral types in the survey. The bi-modal distribution
is largely the product of search techniques; 
the L dwarf sample was identified from 2MASS and DENIS data.}
\end{figure}

In addition to the common proper motion analysis of faint sources, we
inspect the core of each of the primaries to search for extended emission 
associated with a marginally resolved binary. Second-epoch observations 
are used to obtain evidence of common proper motion and to
mitigate systematic psf-distortion effects due to errors in 
phasing of the segmented primary mirror. Point-like sources 
observed nearby in the sky and within an hour of the target observations 
serve as psf measurements. Dithered images of candidate 
binaries and psf stars are not shifted and combined but are treated as 
independent data sets. Psf stars are fit in duplicate to each of the 
candidate binary images using a least-squares minimization method, to
determine component properties.

\subsection{HST/WFPC2 Imaging Survey}

High contrast companions within 0.5$''$ are better detected at spatial
resolution that is not hampered by 
the effects of atmospheric seeing. We are carrying out a companion
program with the Wide Field Planetary Camera 2 on HST to detect close
companions to low-luminosity dwarfs. Equal-luminosity components 
are resolved at 0.09$''$ and contrasting luminosities of $\delta$m = 5 
(I band) are detectable outside 0.32$''$ from the star.

\section{Preliminary Results - Abundant L-Dwarf Binaries?}

In preliminary analysis of Keck/NIRC image
frames for which dual-epoch observations have been obtained, three objects met 
our criteria for reliable identification of a true close binary system
(Koerner et al.\ 1999), including one imaged previously with HST/NICMOS by
Mart\'{\i}n et al.\ (1999).
Contour plots of three L-dwarf binaries are displayed in Fig.\ 2, together 
with the psf stars used decompose them into separate components.
In Fig.\ 3 are plotted the results of psf-fits to obtain the 
separation and PA for the components of DENIS-P J1228,
DENIS-P J0205, and 2MASSW J1146. Mean values
are $0.27\pm0.03''$, $0.51\pm0.03''$, 
$0.29\pm0.06''$ and $33\pm15^\circ$, 
$92\pm18^\circ$, $206\pm19^\circ$, respectively. Projected separations 
correspond to physical separations of 4.9, 9.2, and 7.6 AU at distances 
implied by obtained trigonometric parallaxes (Dahn et al. 2000).
Flux-component ratios for the binaries are $1.1\pm0.4$,  
$1.0\pm0.4$, and $1.0\pm0.3$, respectively. 

\begin{figure}
\plotfiddle{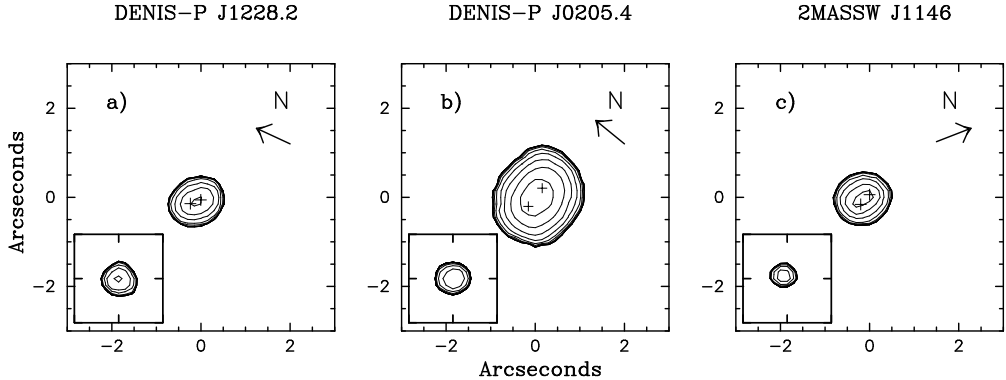}{0pt}{270}{50} {50} {-200}{50}
\vskip 2.0truein
\caption{ {\bf a)} Contour plot taken from Koerner et al.\ (1999)
of K-band imaging of DENIS-P J1228 together with
that of Kelu-1, the ``psf'' star used to derive binary component parameters. 
Contours are at logarithmic intervals. Crosses mark the separation and PA of
components derived in the psf fits to the data shown here. 
{\bf b)} Plots as in a) for DENIS-P J0205 and
associated psf star, LP 647-13, {\bf c)} Plots as in a) and b) for 2MASSW J1146
and psf star 2MASSW J1145.}
\end{figure}

The binary systems presented here have similar projected separations
(5 to 9 AU) and luminosity ratios near unity. They represent the first
binary detections in preliminary analysis of a larger dual-epoch 
survey in which only 10 L-dwarf images have been completely analyzed 
in two epochs. No companions with wider separations or more highly
contrasting luminosities were found thus far. 
These preliminary results suggest a
conjecture for further testing: namely, that multiple systems are 
common in the L dwarf population, that their distribution peaks at radial
separations like that of both Jovian planets in our solar system and M 
dwarfs generally ($\sim5-30$ AU), and that low-contrast mass ratios are 
common. The latter claim is especially in need of testing, since
our survey is not very sensitive to companions at the separations 
reported here if they have high luminosity contrast ratios. Further,
the magnitude-limited surveys from which our sample is taken are biased
toward the detection of equal-luminosity binaries,
since their combined luminosity is greater than for single stars of the same
spectral type.  Ultimately, techniques with both high resolution and high 
dynamic range must be applied to a larger sample to reliably
identify the distribution of circumstellar bodies that encircles this 
population of very cool objects.

\begin{figure}
\plotfiddle{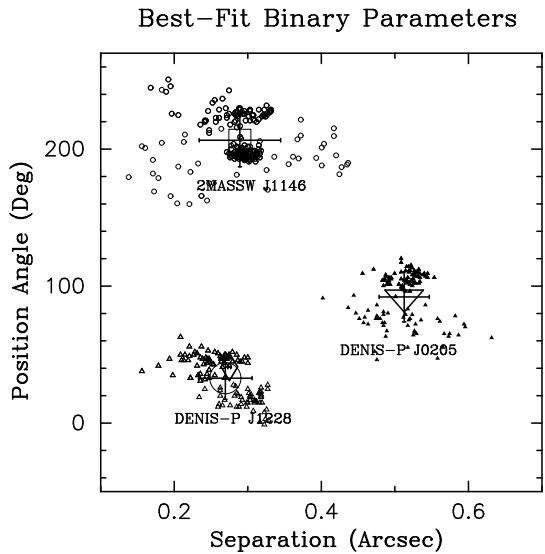}{0pt}{270}{40} {40} {-130}{25}
\vskip 2.75truein
\caption{Binary separation and position angle from psf-fits 
of individual frames for 2MASSW J1146 (small open circles), DENIS-P J1228
(small open triangles), and DENIS-P J0205 (small filled triangles). The mean
of each of the measurements is plotted as a large open symbol for each
object with error bars that mark the rms deviation about the mean. The
HST/NICMOS result for DENIS-P J1228 is plotted as an open diamond
(from Koerner et al.\ 1999).}
\end{figure}

\section{Bayesian Inference of the Underlying Companion Distribution}

We would like especially to discover the probability distribution of 
stellar and planetary companions in order to constrain theories
of their origin; a successful theory should account for that distribution.
In addition, an intensely human interest drives us to seek to 
understand how many
habitable circumstellar environments exist and how typical is the planet
on which we find ourselves.
It will be decades at least before the inventory of circumstellar
objects is complete enough to rely on counting statistics alone to 
provide the whole answer. In the interim, some regions of parameter space for
model probability distributions will be more completely sampled than others.
Relatively luminous companions at distances of 100 AU will soon be 
largely accounted for in nearly all stars detected nearby, for example. 
As parts of this census come to light, it will be challenging to ascertain 
the reliability of distribution estimates for substellar companions
that are based on counting in incomplete samples.

Strictly speaking, the sampling of the companion distribution in our 
completed and combined Keck and HST surveys will still be incomplete, since 
high-luminosity-contrast companions close to the star or wide companions 
separated by more than 20$"$ could go undetected. Furthermore, the range of 
linear separations is inhomogeneously sampled, since a wide range of
distances is represented by the source list. Rather than simply count the 
number of detections and correct for incompleteness, we prefer to use a 
Bayesian model-fitting approach to quantify how well model companion 
distributions are constrained by our data. As outlined below, this approach 
yields the relative probability of a model distribution, given the data. By 
calculating this for a suitable range of models, we can determine both the 
most likely model, and the degree to which this choice is mandated by the 
data.

According to Bayes Theorem, the probability of a model given the data, 
$P<M|D>$,
is calculated by multiplying the probability of the data given the model, 
$P<D|M>$, times the {\it a priori} probability of the model, $P<M>$. For the 
case of a model distribution that is a function only of linear
separation, $R = \theta/\pi$ with angular separation $\theta$ and 
trigonometric parallax $\pi$, and luminosity $L$, 
 the calculation of $M(R,L)$ is straightforward for
an individual image frame $D_i$. The probability of a null
detection is simply one minus the probability of a detection. Since the 
model is, {\it itself,} the probability of a detection, we can
calculate this by integrating over one minus  $M(R,L)$ as simply 
$$ P<D_i|M> \ = {\int^{R_{out}}_{R_{in}}}{\int^{L_{prim}}_{L_{ul}}} (1 - M(R,L)) dR dL $$
\noindent
where $R_{in}$ and $R_{out}$ define the inner and outer linear separations to
which the image is sensitive, and $L_{prim}$ and $L_{ul}$ are, respectively,
the luminosity of the primary and the upper-limit luminosity for the detection
of a companion. Typically, $L_{ul}$ = $L_{ul}(R)$ for small angular
separations. For images where a companion is detected at separation $R'$
with luminosity $L'$, the probability of the result, given the data, is given
by $$ P<D_i|M> \ = {\int_{R{in}}^{R_{out}}}{\int^{L_{prim}}_{L_{ul}}}
\delta(R',L') M(R,L) dR dL$$ 
\noindent
where $\delta(R',L')$ is the Dirac delta function. 
These terms may then be multiplied by the prior probability of the model
according to Bayes Theorem. In the absence of any previous notions about the
distribution, a ``flat prior'' may used by simply setting $P<M>$ = 1.

The probability of a particular model, given all the image frames,
is then the normalized sum of the probabilities for the N individual frames:
$$ P<M|D> \ = {{\sum {P<M|D_i>} } \over N }\ \ .$$

\noindent
A wide range of models may be compared in this way by calculating
the relative probability,
$P_{rel}<M_j|D>$, for the $j^{th}$ model and normalizing over the whole
suite of models considered:

$$ P_{rel}<M_j|D> \ = {P<M_j|D> \over {\sum P<M_i|D>} }\ \ .$$

This approach has the advantage of bringing to bear all the information
inherent in an inhomogeneous data set and weighting proportionally
its influence on the choice of the 
most probable models. If, for example, only a 
few images test
the model in some range of linear separations, their contribution
to the overall probability will be small, such that models which vary 
in their estimate of companions at those separations will not have
widely contrasting probabilities. Conversely,
constraints will be strong in regions of the model parameter space
that are densely
sampled by the data. By considering an appropriate
range of models, confidence levels as a function of parameter values can 
be attached to the best-fit model.

\section{Parametrizing the Model Distribution}

The scientific usefulness of the above methodology
will depend heavily on the choice of models considered. It is 
possible, of course, to aim only to fit some analytic function to the
data so as to derive a best-fit simplified representation of
the observations. This
is done easily by simply fitting a model which is parametrized
in the directly measurable quantities, angular separation and
relative luminosity. But it would be more worthwhile to derive a
distribution function with physically meaningful parameters
that have theoretical significance. The underlying properties 
which describe binary systems most {\it completely} are the orbital
elements and masses. But theories of origin may be constrained by
a few of these or by derivative quantities, such as semi-major axes
or angular momentum. We note, for example, that binary origin
simulations show a marked dependence on $\beta$,
the ratio of rotational to gravitational energy
in the original cloud (cf. Bonnell \& Bastien 1992).

The testing of underlying physical models can proceed as above, so
long as an appropriate transformation exists between the physical
quantity and what is observed. For example, the most probable value
of the semi-major axis, $a_{\rm rel}$, for a companion with observed
angular separation $\theta$ and system distance $d$
has been estimated by Fischer \& Marcy (1992) using Monte Carlo simulations 
to be $$<a_{\rm rel}>\ = 1.26d<\theta> \ \ .$$
\noindent
This transformation can be incorporated easily into a scheme to determine
an underlying model distribution of companions which is a function 
of $<a_{\rm rel}>$ rather than $\theta$. For main sequence stars, a 
transformation between luminosity and mass can be accomplished with 
relationships obtained by dynamic mass determinations (cf. Henry et al.\ 1999).
For L dwarfs, the situation is not well-determined empirically
but requires theoretical models which relate mass, age, and
luminosity (e.g., Burrows et al.\ 1997). To obtain the coveted distribution 
of masses from these relations, 
assumptions about stellar ages will be required. 

Further complications are introduced by the inclusion of higher-order 
multiple systems and, ultimately, in the consideration of planetary systems
as well. The increased effort may well be worth the undertaking, since it 
may yield a general taxonomic classification of multiple dwarf 
and planetary systems with theoretical significance and descriptive
power for characterizing the frequency and types of circumstellar systems.
To this end, we will fit models in a variety of parametrized prescriptions.
We thus consider the application of Bayesian inference to the problem of
the low-luminosity dwarf companion frequency to comprise a pilot study
for larger objectives.


\begin{references}

\reference Basri, G., \& Mart\'{\i}n, E.L. 1997, in ASP Conf. Ser. 134, 
Brown Dwarfs and Extrasolar Planets, ed. R. Rebolo, E. Martin, \& M.R. 
Zapatero-Osorio (San Francisco: ASP), 284

\reference Becklin, E.E., \& Zuckerman, B. 1988, Nature, 336, 656

\reference Black, D.C., 1997, \apj 490, L171

\reference Bonell, I., \& Bastien, P. 1992, \apj 401, 654

\reference Burrows, A., Marley, M., Hubbard, W.B., Lunine, J.I., Guillot, T.,
Saumon, D., Freedman, R., Sudarsky, D., \& Sharp, C. 1997, \apj 491, 856


\reference Dahn, C., et al, 1999, in preparation

\reference Fischer, D.A., \& Marcy, G.W. 1992, \apj, 396 


\reference Henry, T.J., \& McCarthy, D.W.Jr. 1993, \aj, 106, 773

\reference Henry, T.J., Franz, O.G., Wasserman, L.H., Benedict, G.F.,
Shelus, P.J., Ianna, P.A., Kirkpatrick, J.D., \& McCarthy, D.W.Jr., 1999, 
\apj, 512, 864

\reference Kirkpatrick, J.D., Reid, I.N., Leibert, J., Cutri, R.M.,
Nelson, B., Beichman, C., Dahn, C., Monet, D.G., Gizis, J.E., \&
Skrutskie, M.F. 1999, \apj, 519, 802

\reference Kirkpatrick, J.D., et al. 2000, in preparation

\reference Koerner, D.W., Kirkpatrick, J.D., McElwain, M.W., 
\& Bonaventura, N.R. 1999, \apj, in press


\reference Marcy, G.W., \& Butler, R.P. 1996, \apj, 464, L147 

\reference Marcy, G.W., Butler, R.P., Vogt, S.S., Fischer, D.,
Lissauer, J.J. 1998, \apj, 505, L147

\reference  Mart\'{\i}n, E.L., Basri, G., Delfosse, X., \& Forveille, t.,
1997, \aap, 327, L29

\reference  Mart\'{\i}n, E.L., Brandner, W., \& Basri, G. 1999,
Science, 283, 1718

\reference Mayor, M., \& Queloz, D, 1995, Nature, 378, 355

\reference
Nakajima, T., Oppenheimer, B.R., Kulkarni, S.R., Golimowski, D.A., Matthews,
K., \& Durrance, T. 1995, Nature, 378, 463 

\reference  Rebolo, R., Zapatero-Osorio, M.R., Madruga, S.,
Bejar, V.J.S., Arribas, S., \& Licandro, J. 1998, Science, 282, 1309

\reference Reid, I.N., \& Gizis, J.E. 1997, \aj, 113, 2246

\reference Reid, I.N., Kirkpatrick, J.D., Liebert, J., Burrows, A.,
Gizis, J.E., Burgasser, A., Dahn, C.C., Monet, D., Cutri, R., Beichman, C.A.,
Skrutskie, M. 1999, \apj, 521, 613





\end{references}
\end{document}